\documentclass[submission,copyright,creativecommons]{eptcs}
\providecommand{\event}{ACL2 Workshop 2023}
\usepackage{underscore, latexsym,pifont,amsmath,amstext,amssymb,amsfonts,makeidx,url}
\title{A Formalization of Finite Group Theory: Part II}
\author{David M.~Russinoff
\email{david@russinoff.com}
}
\def\titlerunning{A Formalization of Finite Group Theory: Part II}
\def\authorrunning{D.M. Russinoff}

\begin{document}
\maketitle

\begin{abstract}
This is the second installment of an exposition of an ACL2 formalization of finite group theory.  The first, which was presented at the 2022 ACL2 workshop, covered groups and subgroups, cosets, normal subgroups, and quotient groups, culminating in a proof of Cauchy's Theorem: {\it If the order of a group $G$ is divisible by a prime $p$, then $G$ has an element of order $p$.}  This sequel addresses homomorphisms, direct products, and the Fundamental Theorem of Finite Abelian Groups: {\it Every finite abelian group is isomorphic to the direct product of a list of cyclic p-groups, the orders of which are unique up to permutation.}   This theorem is a suitable application of ACL2 because of its extensive reliance on recursion and induction as well as the constructive nature of the factorization.  The proof of uniqueness is especially challenging, requiring the formalization of vague intuition that is commonly taken as self-evident.
\end{abstract}

\section{Introduction}

In comparison to higher-order logic theorem provers, ACL2 offers a high degree of proof automation at the expense of logical expressiveness.  With regard to the formalization of pure mathematics,  basic concepts are often difficult to formulate in a first-order logic, but when this obstacle is overcome, proofs are relatively straightforward.  This prospect has motivated our pursuit of an ACL2 formalization of abstract algebra, beginning with finite group theory, an area in which substantial progress has already been achieved with other provers, most notably the Coq proof assistant \cite{gonthier}.  Our investigation of this subject, which deals with properties of operations defined on sets, must address two limitations of the ACL2 logic: (1)~quantification over functions is not provided, and (2) ACL2 data are ordered lists rather than sets.  A common solution to these problems is the use of constrained functions.  In particular, a natural approach to the formalization of group theory begins with an {\tt encapsulate} form that introduces a set of constrained functions including a predicate representing group membership, a binary group operation, and a unary inverse operator.  An alternative scheme based on {\tt defn-sk} was devised by Yuan Yu in his 1990 Nqthm formalization \cite{yu}, which included a proof of Lagrange's Theorem: {\it The order of a group is divisible by that of any subgroup.}

Our original submission on this subject to the 2022 ACL2 Workshop \cite{workshop} was motivated by the observation that any significant progress beyond Lagrange's Theorem would require the facility of proof by induction on the order of a group, which is apparently unavailable through either of the methods mentioned above.  That is, a more productive ACL2 formalization of group theory would begin with an explicit predicate that recognizes groups of arbitrary well-defined orders.  (This approach necessarily limits the investigation to finite groups.)  Thus, the predicate {\tt groupp} defines a group of order $n$ to be an $n\times n$ matrix (a list of $n$ rows of length $n$) representing the group's operation table, which is stipulated to satisfy the usual group axioms.  The first row of the matrix is the list of group elements, the order of which is insignificant except that the first element must be the identity:

\begin{small}
\begin{verbatim}
  (defmacro elts (g) `(car ,g))
  (defmacro in (x g) `(member-equal ,x (elts ,g)))
  (defund e (g) (caar g))
\end{verbatim}
\end{small}
The index of an element {\tt x} of {\tt g}, {\tt (ind x g)}, is its position in the element list:
\begin{small}
\begin{verbatim}
  (defun index (x l)
    (if (consp l)
        (if (equal x (car l))
            0
          (1+ (index x (cdr l))))
      0))
  (defmacro ind (x g) `(index ,x (elts ,g)))
\end{verbatim}
\end{small}
Thus, the group operation is defined by
\begin{small}
\begin{verbatim}
  (defund op (x y g)
    (nth (ind y g)
         (nth (ind x g) g)))
\end{verbatim}
\end{small}
The lack of ACL2 support for unordered sets is mitigated by exploiting the notion of an ordered list of elements of a group.  Thus, for example, a left coset of a subgroup is defined to be ordered with respect to the larger group, thereby ensuring that intersecting cosets are equal.  Note that a subgroup need not be ordered with respect to its parent.  For example, a cyclic subgroup has a natural ordering that is generally different from that of the parent group.  In most cases, however, we arrange for the ordering to be inherited.

In order to circumvent the cumbersome explicit construction of the defining table for every group of interest, we introduce a {\tt defgroup} macro that generates a parametrized group definition and proofs of the axioms (through functional instantiation) once the user has supplied the element list and terms specifying the binary operation and the inverse operator.  This is used, for example, in the definition of the quotient group of a normal subgroup as well as the symmetric groups.  A similar {\tt defsubgroup} macro facilitates the definition of parametrized subgroups, e.g., the centralizer of a group element, the center of a group, cyclic subgroups, and intersections of subgroups.  As a proof of concept, the culmination of the 2022 submission is an inductive proof of a theorem of Cauchy: {\it If the order of a group $G$ is divisible by a prime $p$, then $G$ has an element of order $p$.}  The theory up to this point is embodied in the first four books of the ACL2 directory {\tt books/projects/groups}: {\tt lists}, {\tt groups}, {\tt quotients}, and {\tt cauchy}.

The 2022 paper is a prerequisite for a reading of this sequel, which presents a continuation of the theory comprising three additional books of the {\tt groups} directory.  The first of these, {\tt maps}, addresses another class of functions that must be encoded in the logic: group homomorphisms (Section~\ref{maps}), which we represent as association lists.  The second, {\tt products}, covers direct products (Section~\ref{products}).  The results of these two books are applied in the third, {\tt abelian}, in a proof of the Fundamental Theorem of Finite Abelian Groups: {\it Every finite abelian group is isomorphic to the direct product of a list of cyclic p-groups, the orders of which are unique up to permutation.}  The proof also applies various number-theoretic results from the book {\tt projects/numbers/euclid}.  The proof is in three parts: (1)~the factorization of an abelian p-group as a product of cyclic groups (Section~\ref{pgroups}, (2)~the factorization of an arbitrary abelian group as a product of p-groups (Section~\ref{abelian}), and (3)~the uniqueness of the factorization (Section~\ref{uniqueness}).  This theorem is a suitable application of ACL2 because of its extensive reliance on recursion and induction as well as the constructive nature of the factorization.  The proof of uniqueness is especially challenging, requiring the formalization of vague intuition that is usually taken as self-evident \cite{rotman}.

\section{Homomorphisms}\label{maps}

A {\it homomorphism} is a function {\tt f} from a group {\tt g} to a group {\tt h} that preserves the group operation, i.e., satisfies the identity {\tt (op x y g)} = {\tt (op (f x) (f y) h)}, and thus relates the algebraic properties of {\tt g} and {\tt h}.  Of particular interest is the case of a bijective homomorphism, or an {\it isomorphism}, which establishes the algebraic equivalence of two groups.

In order to introduce group homomorphisms into our theory, we define a {\it map} to be an alist with pairwise distinct cars:
\begin{small}
\begin{verbatim}
  (defun cons-listp (m) 
    (if (consp m) (and (consp (car m)) (cons-listp (cdr m))) (null m)))
  (defund domain (m) (strip-cars m))
  (defund mapp (m) (and (cons-listp m) (dlistp (domain m))))
\end{verbatim}
\end{small}
The function {\tt mapply} applies a map to an element of its domain:
\begin{small}
\begin{verbatim}
  (defund mapply (map x) (cdr (assoc-equal x map)))
\end{verbatim}
\end{small}
The macro {\tt defmap} provides a convenient method of defining maps.  The macro call
\begin{center}
  {\tt (defmap} {\it name args domain val}{\tt )}
\end{center}
defines a family of maps parametrized by {\it args} with given {\it domain}, which is assumed to be a dlist.  The parameter {\it val} is the value that the map assigns to an element x of its domain.  For example, the following form automates the construction of a composition of two maps:
\begin{small}
\begin{verbatim}
  (defmap compose-maps (map2 map1)
    (domain map1)
    (mapply map2 (mapply map1 x)))
\end{verbatim}
\end{small}
Evaluation of this form generates two definitions and proves three lemmas:
\begin{small}
\begin{verbatim}
  (DEFUN COMPOSE-MAPS-AUX (L MAP2 MAP1)
    (IF (CONSP L)
        (LET ((X (CAR L)))
          (CONS (CONS X (MAPPLY MAP2 (MAPPLY MAP1 X)))
                (COMPOSE-MAPS-AUX (CDR L) MAP2 MAP1)))
      NIL))
  (DEFUN COMPOSE-MAPS (MAP2 MAP1)
    (COMPOSE-MAPS-AUX (DOMAIN MAP1) MAP2 MAP1))
  (DEFTHM DOMAIN-COMPOSE-MAPS
    (IMPLIES (DLISTP (DOMAIN MAP1))
             (EQUAL (DOMAIN (COMPOSE-MAPS MAP2 MAP1))
                    (DOMAIN MAP1))))
  (DEFTHM MAPP-COMPOSE-MAPS
    (IMPLIES (DLISTP (DOMAIN MAP1))
             (MAPP (COMPOSE-MAPS MAP2 MAP1))))
  (DEFTHM COMPOSE-MAPS-VAL
    (IMPLIES (MEMBER-EQUAL X (DOMAIN MAP1))
             (EQUAL (MAPPLY (COMPOSE-MAPS MAP2 MAP1) X)
                    (MAPPLY MAP2 (MAPPLY MAP1 X)))))
\end{verbatim}
\end{small}
A {\it homomorphism} from a group {\tt g} to a group {\tt h} is a map that satisfies the following predicate:
\begin{small}
\begin{verbatim}
  (defund homomorphismp (map g h)
    (and (groupp g)
         (groupp h)
         (mapp map)
         (sublistp (elts g) (domain map))
         (equal (mapply map (e g)) (e h))
         (not (codomain-cex map g h))
         (not (homomorphism-cex map g h))))
\end{verbatim}
\end{small}
The functions {\tt codomain-cex} and {\tt homomorphism-cex} search for counterexamples of these two properties:
\begin{small}
\begin{verbatim}
  (implies (in x g)
           (in (mapply map x) h))
  (implies (and (in x g) (in y g))
           (equal (mapply map (op x y g))
                  (op (mapply map x) (mapply map y) h)))
\end{verbatim}
\end{small}
Once these two implications have been proved to hold for a proposed homomorphism, the corresponding conjuncts of the definition are readily established.  In this manner, it is easily shown, for example, that a composition of homomorphisms is a homomorphism:
\begin{small}
\begin{verbatim}
  (defthm homomorphismp-compose-maps
    (implies (and (homomorphismp map1 g h) (homomorphismp map2 h k))
             (homomorphismp (compose-maps map2 map1) g k)))
\end{verbatim}
\end{small}

The {\it image} of a homomorphism {\tt map} from {\tt g} to {\tt h} is a subgroup of {\tt h}.  Its element list is defined using the function {\tt insert}, which ensures that it is ordered with respect to {\tt h}:
\begin{small}
\begin{verbatim}
  (defun ielts-aux (map l h)
    (if (consp l)
        (insert (mapply map (car l))
                (ielts-aux map (cdr l) h)
                h)
      ()))
  (defund ielts (map g h)
    (ielts-aux map (elts g) h))
  (defthm ordp-ielts
    (implies (homomorphismp map g h)
             (ordp (ielts map g h) h)))
\end{verbatim}
\end{small}
The group {\tt (image g h)} is automatically defined by {\tt defsubgroup} once the usual prerequisite lemmas have been proved:
\begin{small}
\begin{verbatim}
  (defthm dlistp-ielts
    (implies (homomorphismp map g h)
             (dlistp (ielts map g h))))
  (defthm sublistp-ielts
    (implies (homomorphismp map g h)
             (sublistp (ielts map g h) (elts h))))
  (defthm consp-ielts
    (implies (groupp g)
             (consp (ielts map g h))))
  (defthm ielts-closed
    (implies (and (homomorphismp map g h)
                  (member-equal x (ielts map g h))
                  (member-equal y (ielts map g h)))
             (member-equal (op x y h) (ielts map g h))))
  (defthm ielts-inverse
    (implies (and (homomorphismp map g h)
                  (member-equal x (ielts map g h)))
             (member-equal (inv x h) (ielts map g h))))
\end{verbatim}
\end{small}
The arguments of {\tt defsubgroup} are the subgroup parameters, the parent group, the constraint that must be satisfied by the parameters, and the element list:
\begin{small}
\begin{verbatim}
  (defsubgroup image (map g) h
    (homomorphismp map g h)
    (ielts map g h))
\end{verbatim}
\end{small}\medskip

The {\it kernel} of {\tt map} is a subgroup of {\tt g} consisting of those elements that are mapped to the identity element of {\tt h}:
\begin{small}
\begin{verbatim}
  (defun kelts-aux (map l h)
    (if (consp l)
        (if (equal (mapply map (car l)) (e h))
            (cons (car l) (kelts-aux map (cdr l) h))
          (kelts-aux map (cdr l) h))
    ()))
  (defund kelts (map g h)
    (kelts-aux map (elts g) h))
\end{verbatim}
\end{small}
Once again, we invoke {\tt defsubgroup} after establishing its prerequisite lemmas:
\begin{small}
\begin{verbatim}
  (defsubgroup kernel (map h) g
    (homomorphismp map g h)
    (kelts map g h))
\end{verbatim}
\end{small}
The usual classes of homomorphisms are defined in terms of the image and the kernel:
\begin{small}
\begin{verbatim}
  (defund epimorphismp (map g h)
    (and (homomorphismp map g h)
         (equal (image map g h) h)))
  (defund endomorphismp (map g h)
    (and (homomorphismp map g h)
         (equal (kernel map h g) (trivial-subgroup g))))
  (defund isomorphismp (map g h)
    (and (epimorphismp map g h) (endomorphismp map g h)))
\end{verbatim}
\end{small}
The inverse of an isomorphism is defined by {\tt defmap}:
\begin{small}
\begin{verbatim}
  (defun preimage-aux (x map l)
    (if (consp l)
        (if (equal x (mapply map (car l)))
            (car l)
          (preimage-aux x map (cdr l)))
      ()))
  (defund preimage (x map g)
    (preimage-aux x map (elts g)))
  (defmap inv-isomorphism (map g h) (elts h) (preimage x map g))
  (defthmd isomorphismp-inv
    (implies (isomorphismp map g h)
             (isomorphismp (inv-isomorphism map g h) h g)))
\end{verbatim}
\end{small}
We shall also require the following important property of isomorphisms:
\begin{small}
\begin{verbatim}
  (defthm isomorphismp-compose-maps
    (implies (and (isomorphismp map1 g h) (isomorphismp map2 h k))
                  (isomorphismp (compose-maps map2 map1) g k)))
\end{verbatim}
\end{small}

\section{Direct Products}\label{products}

{\it Direct products} provide a means of constructing complex groups from simpler ones.  Given a non-null proper list of groups {\tt l}, we shall define the group {\tt (direct-product l)}.  Its element list is the Cartesian product {\tt (group-tuples l)}, defined as follows:
\begin{small}
\begin{verbatim}
  (defun conses (x l)
    (if (consp l)
        (cons (cons x (car l)) (conses x (cdr l)))
      ()))
  (defun group-tuples-aux (l m)
    (if (consp l)
        (append (conses (car l) m)
                (group-tuples-aux (cdr l) m))
      ()))
  (defun group-tuples (l)
    (if (consp l)
        (group-tuples-aux (elts (car l)) (group-tuples (cdr l)))
      (list ())))
\end{verbatim}
\end{small}
It is easily shown that the length of {\tt (group-tuples l)} is the product of the orders of the members of {\tt l}.  If {\tt x} is a member of this list, then {\tt x} is a list of the same length as {\tt l}, and each member of {\tt x} is a group element of the corresponding member of {\tt l}.  The {\tt car} of {\tt (group-tuples l)}, which will be the identity element of the direct product, is the list defined as follows:
\begin{small}
\begin{verbatim}
  (defun e-list (l)
    (if (consp l)
        (cons (e (car l)) (e-list (cdr l)))
      ()))
\end{verbatim}
\end{small}
The group operation and inverse operator are defined recursively:
\begin{small}
\begin{verbatim}
  (defun dp-op (x y l)
    (if (consp l)
        (cons (op (car x) (car y) (car l))
              (dp-op (cdr x) (cdr y) (cdr l)))
      ()))
  (defun dp-inv (x l)
    (if (consp l)
        (cons (inv (car x) (car l))
              (dp-inv (cdr x) (cdr l)))
      ()))
\end{verbatim}
\end{small}
Once the requisite group properties are proven, we invoke {\tt defgroup} to construct the direct product:
\begin{small}
\begin{verbatim}
  (defgroup direct-product (l)
    (and (group-list-p l) (consp l)) ;parameter constraint
    (group-tuples l)                 ;element list
    (dp-op x y l)                    ;group operation
    (dp-inv x gl))                   ;inverse operator
\end{verbatim}
\end{small}
The ordering of the elements of the direct product is given by the following, which is useful in proving that a given subgroup is ordered:
\begin{small}
\begin{verbatim}
  (defthmd ind-dp-compare
    (implies (and (group-list-p l)
                  (consp l)
                  (in x (direct-product l))
                  (in y (direct-product l)))
             (iff (< (ind x (direct-product l))
                     (ind y (direct-product l)))
                  (or (< (ind (car x) (car l))
                         (ind (car y) (car l)))
                      (and (consp (cdr l))
                           (equal (car x) (car y))
                           (< (ind (cdr x) (direct-product (cdr l)))
                              (ind (cdr y) (direct-product (cdr l)))))))))
\end{verbatim}
\end{small}

A construction related to the direct product is the list of products of elements of two subgroups {\tt h} and {\tt k} of a group {\tt g}.  Note that the definition ensures that this list is ordered with respect to {\tt g}:
\begin{small}
\begin{verbatim}
  (defun products-aux (l g)
    (if (consp l)
        (insert (op (caar l) (cadar l) g) (products-aux (cdr l) g) g)
      ()))
  (defund products (h k g)
    (products-aux (group-tuples (list h k)) g))
\end{verbatim}
\end{small}
While {\tt (products h k g)} does not in general form a subgroup of {\tt g}, it does when either {\tt h} or {\tt k} is normal:
\begin{small}
\begin{verbatim}
  (defsubgroup product-group (h k) g
    (and (subgroupp h g)
         (subgroupp k g)
         (or (normalp h g) (normalp k g)))
    (products h k g))
\end{verbatim}
\end{small}
When {\tt h} and {\tt k} are both normal in {\tt g}, so is {\tt (product-group (h k g)}.

We have the following formula for the length of {\tt (products h k g)}:
\begin{small}
\begin{verbatim}
  (defthmd len-products
    (implies (and (subgroupp h g)
                  (subgroupp k g))
             (equal (len (products h k g))
                    (/ (* (order h) (order k))
                       (order (group-intersection h k g))))))
\end{verbatim}
\end{small}
The derivation of this formula is based on the following function, which converts a list {\tt l} of members of {\tt (lcosets (group-intersection h k g) h)} to a list of members of (lcosets k g) by replacing each member {\tt c} of {\tt l} with {\tt (lcoset (car c) k g)}:
\begin{small}
\begin{verbatim}
  (defun lift-cosets-aux (l k g)
    (if (consp l)
        (cons (lcoset (caar l) k g)
              (lift-cosets-aux (cdr l) k g))
      ()))
\end{verbatim}
\end{small}
We apply {\tt lift-cosets-aux} to the full list {\tt (lcosets (group-intersection h k g) g)}:
\begin{small}
\begin{verbatim}
  (defund lift-cosets (h k g)
    (lift-cosets-aux (lcosets (group-intersection h k g) h) k g))
\end{verbatim}
\end{small}
The result is a list of distinct elements of (lcosets k g), and therefore, appending them yields a dlist:
\begin{small}
\begin{verbatim}
  (defthm dlistp-append-list-lift-cosets
    (implies (and (subgroupp h g) (subgroupp k g))
             (dlistp (append-list (lift-cosets h k g)))))
\end{verbatim}
\end{small}
The length of {\tt (lift-cosets h k g)} is that of {\tt (lcosets (group-intersection h k g) h)}, which is the quotient
\begin{center}
{\tt(/ (order h) (order (intersect-groups h k g)))}
\end{center}
Since the length of each member of the list is {\tt (order k)}, we have the following expression for the length of the appended list:
\begin{small}
\begin{verbatim}
  (defthm len-append-list-lift-cosets
    (implies (and (subgroupp h g) (subgroupp k g))
             (equal (len (append-list (lift-cosets h k g)))
                    (/ (* (order h) (order k))
                       (order (group-intersection h k g))))))
\end{verbatim}
\end{small}
It is easy to show that each member of {\tt (lift-cosets h k g)} is a sublist of {\tt (products h k g)}.  On the other hand, if {\tt x} is an element of {\tt (products h k g)}, then {\tt x} = {\tt (op a b g)}, where {\tt a} is in {\tt h} and {\tt b} is in {\tt k}. Some member of {(lcosets (group-intersection h k g) h)} contains {\tt a}, as does the corresponding member of {\tt (lift-cosets h k g)}, which therefore contains {\tt x}.  Thus, {\tt(products h k g)} is a sublist of {\tt (append-list (lift-cosets h k g))}.  Since both lists are dlists and each is a sublist of the other, they have the same length, and the formula {\tt len-products} follows from {\tt len-append-list-lift-cosets}.

The product of a list of subgroups is defined recursively:
\begin{small}
\begin{verbatim}
  (defun product-group-list (l g)
    (if (consp l)
        (product-group (car l) (product-group-list (cdr l) g) g)
      (trivial-subgroup g)))
\end{verbatim}
\end{small}
By induction, if each group in {\tt l} is normal in {\tt g}, then so is {\tt (product-group-list l g)}.  If {\tt l} satisfies the following predicate, then that subgroup is isomorphic to {\tt (direct-product l)}:
\begin{small}
\begin{verbatim}
  (defun internal-direct-product-p (l g)
    (if (consp l)
        (and (internal-direct-product-p (cdr l) g)
             (normalp (car l) g)
             (equal (group-intersection (car l) (product-group-list (cdr l) g) g)
                    (trivial-subgroup g)))
      (null l)))
\end{verbatim}
\end{small}
Moreover, if that subgroup has the same order as {\tt g}, then since it inherits the ordering of {\tt g}, the two groups are equal.  The isomorphism is conveniently constructed by {\tt defmap}:
\begin{small}
\begin{verbatim}
  (defun product-list-val (x g)
    (if (consp x)
        (if (consp (cdr x))
            (op (car x) (product-list-val (cdr x) g) g)
          (car x))
      ()))
  (defmap product-list-map (l g)
    (group-tuples l)
    (product-list-val x g))
  (defthmd isomorphismp-dp-idp
    (implies (and (groupp g)
                  (consp l)
                  (internal-direct-product-p l g)
                  (= (product-orders l) (order g)))
             (isomorphismp (product-list-map l g)
                           (direct-product l)
                           g)))
\end{verbatim}
\end{small}
Note also that the product of two non-intersecting internal direct products is an internal direct product:
\begin{small}
\begin{verbatim}
(defthmd internal-direct-product-append
  (implies (and (internal-direct-product-p l g)
                (internal-direct-product-p m g)
                (equal (group-intersection (product-group-list l g)
                                           (product-group-list m g)
                                           g)
                       (trivial-subgroup g)))                
           (internal-direct-product-p (append l m) g)))
\end{verbatim}
\end{small}

\section{Factorization of p-Groups}\label{pgroups}

A group is {\it abelian} if its operation is commutative.  The notion of an abelian group may be viewed as an abstraction of the familiar numerical groups: the additive groups of integers, modular integers, rationals, and reals, and the multiplicative groups of the non-zero rationals and reals.  Finite abelian groups admit a particularly simple classification as direct products of cyclic groups.  We begin with the case of an abelian {\tt p-group}, the order of which is a power of a prime $p$:
\begin{small}
\begin{verbatim}
  (defund p-groupp (g p)
    (and (primep p) (groupp g) (powerp (order g) p)))
\end{verbatim}
\end{small}
In this section, we shall prove that every abelian $p$-group is an internal direct product of cyclic subgroups.  This will follow by induction once we show that if such a group is not cyclic, then it is an internal direct product of two non-trivial subgroups.  The proof of this result is also inductive, based on the notion of ``lifting'' a subgroup of a quotient group.  If {\tt n} is a normal subgroup of {\tt g} and {\tt h} is a subgroup of {\tt (quotient g n)}, then {\tt (lift h n g)} is the subgroup of {\tt g} formed by appending the cosets that belong to {\tt h}.  In the book {\tt groups/quotients}, we prove the following two lemmas:
\begin{small}
\begin{verbatim}
  (defthmd lift-subgroup
    (implies (and (normalp n g) (subgroupp h (quotient g n)))
             (subgroupp n (lift h n g))))

  (defthmd lift-order
    (implies (and (normalp n g) (subgroupp h (quotient g n)))
             (equal (order (lift h n g)) (* (order h) (order n)))))
\end{verbatim}
\end{small}
Assume {\tt (p-groupp g p)} and {\tt (abelianp g)}.  Our objective is to construct subgroups {\tt g1} and {\tt g2} of {\tt g} that satisfy the following predicate:
\begin{small}
\begin{verbatim}
  (defund desired-properties (g g1 g2)
    (and (subgroupp g1 g)
         (cyclicp g1)
         (subgroupp g2 g)
         (equal (* (order g1) (order g2)) (order g))
         (equal (group-intersection g1 g2 g) (trivial-subgroup g))))
\end{verbatim}
\end{small}
The construction begins with the selection of an element {\tt a} of maximal ord in {\tt g}, as computed by the function {\tt max-ord}:
\begin{small}
\begin{verbatim}
  (defun max-ord-aux (g n)
    (if (zp n)
        1
      (if (elt-of-ord n g)
          n
        (max-ord-aux g (1- n)))))
  (defund max-ord (g)
    (max-ord-aux g (order g)))
\end{verbatim}
\end{small}
For convenience, we collect these hypotheses in another predicate:
\begin{small}
\begin{verbatim}
  (defund phyp (a p g)
    (and (p-groupp g p)
         (abelianp g)
         (not (cyclicp g))
         (in a g)
         (equal (ord a g) (max-ord g))))
\end{verbatim}
\end{small}
The first of the two subgroups is the cyclic group generated by {\tt a}:
\begin{small}
\begin{verbatim}
  (defund g1 (a g) (cyclic a g))
\end{verbatim}
\end{small}
(We shall abbreviate the term {\tt (g1 a g)} as {\tt g1}; related terms defined below will be similarly abbreviated.)
By {\tt cauchy}, some coset in {\tt (quotient g g1)} has order {\tt p}.  We define {\tt x\$} to be a member of that coset:
\begin{small}
\begin{verbatim}
  (defund x$ (a p g) (car (elt-of-ord p (quotient g (g1 a g)))))
\end{verbatim}
\end{small}
Thus, {\tt x\$} is not in {\tt g1} but {\tt (power x\$ p g)} is in {\tt g1}.  This implies {\tt (power x\$ p g)} is a member of {\tt (powers a g)}, and hence {\tt (power x\$ p g)} = {\tt (power a i\$ g)}, where {\tt i\$} is defined by
\begin{small}
\begin{verbatim}
  (defund i$ (a p g) (index (power (x$ a p g) p g) (powers a g)))
\end{verbatim}
\end{small}
It follows that {\tt i\$} is divisible by {\tt p}, for otherwise {\tt (power a i\$ g)} has the same order as {\tt a}, implying that {\tt x\$} has order greater than {\tt (max-ord g)}.  Consider the cyclic subgroup {\tt c\$} of {\tt g} defined as follows:
\begin{small}
\begin{verbatim}
  (defund j$ (a p g) (/ (i$ a p g) p))

  (defund y$ (a p g)
    (op (power (inv a g) (j$ a p g) g)
        (x$ a p g)
        g))

  (defun c$ (a p g) (cyclic (y$ a p g) g))
\end{verbatim}
\end{small}
Note that {\tt y\$} is not in {\tt g1}, but 
\begin{small}
\begin{verbatim}
         (power y$ p g) = (op (power (power (inv a g) j$ g) p g) (power x$ p g) g)
                        = (op (power (inv a g) (* j$ p) g) (power x$ p g) g)
                        = (op (power (inv a g) i$ g) (power x$ p g) g)
                        = (op (inv (power a i$ g) g) (power x$ p g) g)
                        = (op (inv (power x$ p g) g) (power x$ p g) g)
                        = (e g)
\end{verbatim}
\end{small}
Thus, {\tt (order c\$) = (ord y\$ g) = p} and {\tt g1} and {\tt c\$} intersect trivially.  Let {\tt g*} and {\tt a*} be defined as follows:
\begin{small}
\begin{verbatim}
  (defun g* (a p g)
    (quotient g (c$ a p g)))
  (defun a* (a p g)
    (lcoset a (c$ a p g) g))
\end{verbatim}
\end{small}
Then {\tt (ord a* g*) = (max-ord g)}, for otherwise a power of {\tt a} less than {\tt (max-ord g)} would be in {\tt c\$} and therefore equal to {\tt (e g)}.  Thus, {\tt a*} has maximal ord in {\tt g\$}.  If {\tt g*} is cyclic, then its order is {\tt (max-ord g)}, which implies
\begin{small}
\begin{verbatim}
        (order g) = (* (order g*) p} = (* (order g1) (order c$))
\end{verbatim}
\end{small}
and we may define {\tt g2} = {\tt c\$}.  Otherwise we proceed by induction on {\tt (order g}, substituting {\tt g*} and {\tt a*} for {\tt g} and {\tt a}.  Let {\tt g1*} = {\tt (cyclic a* g*)}.  By inductive hypothesis, {\tt g*} is the internal direct product of {\tt g1*} and some {\tt g2*}.  We define {\tt g2} to be {\tt (lift g2* c\$ g)}.  This yields the following recursive definition:
\begin{small}
\begin{verbatim}
  (defund g2 (a p g)
    (declare (xargs :measure (order g)))
    (if (phyp a p g)
        (if (cyclicp (g* a p g))
            (c$ a p g)
          (lift (g2 (a* a p g) p (g* a p g))
                (c$ a p g)
                g))
      ()))
\end{verbatim}
\end{small}
We need only show that {\tt (desired-properties g g1 g2)} follows from {\tt (desired-properties g* g1* g2*)}.  We may assume that {\tt g1} is not cyclic.  We have {\tt (order g1)} = {\tt (max-ord g)} = {\tt (order g1*)}, and by {\tt lift-order}, {\tt (order g2)} = {\tt (* p (order g2*)}.  Thus,
\begin{small}
\begin{verbatim}
    (* (order g1) (order g2)) = (* p (order g1*) (order g2*))
                              = (* p (order g*))
                              = (order g).
\end{verbatim}
\end{small}
Finally, suppose {\tt r} is in both {\tt g1} and {\tt g2}.  Then {\tt (lcoset r c\$ g)} is in both {\tt g1*} and {\tt g2*}, which implies {\tt r} is in {\tt c\$}.  But then {\tt r} is in both {\tt g1} and {\tt c\$}, which implies {\tt r} = {\tt (e g)}.  This completes the induction, and we have
\begin{small}
\begin{verbatim}
  (defthmd factor-p-group
    (implies (phyp a p g)
             (desired-properties g (g1 a g) (g2 a p g))))
\end{verbatim}
\end{small}
This result provides the basis of the factorization of {\tt g}.  The goal is to show that {\tt g} is the internal direct product of a list of subgroups characterized as follows:
\begin{small}
\begin{verbatim}
  (defund cyclic-p-group-p (g)
    (and (cyclicp g)
         (> (order g) 1)
         (p-groupp g (least-prime-divisor (order g)))))
  (defun cyclic-p-group-list-p (l)
    (if (consp l)
        (and (cyclic-p-group-p (car l)) (cyclic-p-group-list-p (cdr l)))
      (null l)))
\end{verbatim}
\end{small}
The list is constructed recursively:
\begin{small}
\begin{verbatim}
  (defun cyclic-p-subgroup-list (p g)
    (declare (xargs :measure (order g)))
    (if (and (p-groupp g p) (abelianp g) (> (order g) 1))
        (if (cyclicp g)
            (list g)
          (let ((a (elt-of-ord (max-ord g) g)))
            (cons (g1 a g) (cyclic-p-subgroup-list p (g2 a p g)))))
      ()))
\end{verbatim}
\end{small}
The desired result follows from {\tt factor-p-group} by induction on {\tt (order g)}:
\begin{small}
\begin{verbatim}
  (defthmd p-group-factorization
    (implies (and (p-groupp g p) (abelianp g) (> (order g) 1))
             (let ((l (cyclic-p-subgroup-list p g)))
               (and (consp l)
                    (cyclic-p-group-list-p l)
                    (internal-direct-product-p l g)
                    (equal (order g) (product-orders l))))))
\end{verbatim}
\end{small}

\section{Factorization of Abelian Groups}\label{abelian}

We shall prove constructively that every finite abelian group is isomorphic to a direct product of cyclic p-groups.  The proof is again inductive, based on {\tt p-group-factorization} and the following lemma: {\it If} {\tt (order g)} {\it is the product of relatively prime integers} {\tt m} {\it and} {\tt n}, {\it then} {\tt g} {\it is the internal direct product of two subgroups of orders} {\tt m} {\it and} {\tt n}.  To derive this result, we define the ordered list of all elements of {\tt g} with order dividing {\tt m}:
\begin{small}
\begin{verbatim}
  (defun elts-of-ord-dividing-aux (m l g)
    (if (consp l)
        (if (divides (ord (car l) g) m)
            (cons (car l) (elts-of-ord-dividing-aux m (cdr l) g))
          (elts-of-ord-dividing-aux m (cdr l) g))
      ()))
  (defund elts-of-ord-dividing (m g)
    (elts-of-ord-dividing-aux m (elts g) g))
\end{verbatim}
\end{small}
If {\tt g} is abelian, then this list forms a subgroup of {\tt g}:
\begin{small}
\begin{verbatim}
  (defsubgroup subgroup-ord-dividing (m) g
    (and (abelianp g) (posp m))
    (elts-of-ord-dividing m g))
\end{verbatim}
\end{small}
Let {\tt h = (subgroup-ord-dividing m g)} and {\tt k = (subgroup-ord-dividing n g)}.  If {\tt x} is in both {\tt h} and {\tt k}, then {\tt (ord x)} divides both {\tt m} and {\tt n}, and by {\tt divides-gcd} of the book {\tt euclid}, {\tt (ord x)} divides {\tt (gcd m n) = 1}, which implies {\tt x = (e g)}.  Thus, {\tt h} and {\tt k} intersect trivially:
\begin{small}
\begin{verbatim}
  (defthmd rel-prime-factors-intersection
    (implies (and (groupp g) (abelianp g)
                  (posp m) (posp n) (= (gcd m n) 1))
             (let ((h (subgroup-ord-dividing m g)) (k (subgroup-ord-dividing n g)))
               (equal (group-intersection h k g) (trivial-subgroup g)))))
\end{verbatim}
\end{small}
By {\tt gcd-linear-combination} (book {\tt euclid}), since {\tt (gcd m n)} = 1, there exist {\tt r} and {\tt s} such that \\
{\tt (+ (* r n) (* s m))} = 1.  Let {\tt x} be in {\tt g}.  Then
\begin{small}
\begin{verbatim}
  x = (power x (+ (* r n) (* s m)) g) = (op (power x (* r n) g) (power x (* s m) g) g).
\end{verbatim}
\end{small}
Since
\begin{small}
\begin{verbatim}
   (power (power x (* r n) g) m g) = (power (power x (* m n) g) r g) = (e g),
\end{verbatim}
\end{small}
{\tt (power x (* r n) g)} is in {\tt m}, and similarly, {\tt (power x (* s m) g)} is in {\tt k}.  Thus, by len-products,
\begin{small}
\begin{verbatim}
    (* m n) = (order g) = (len (products h k g)) = (* (order h) (order k)).
\end{verbatim}
\end{small}
If {\tt p} is a prime dividing {\tt (order h)}, then by {\tt cauchy}, {\tt h} has an element of order {\tt p}, and therefore {\tt p} divides {\tt m}, which implies {\tt p} does not divide {\tt n}.  By {\tt lagrange} and {\tt divides-product-divides-factor} (of the book {\tt euclid}), {\tt (order h)} divides {\tt m}, and therefore {\tt (<= (order h) m)}.  Similarly, {\tt (<= (order k) n)}.  Since {\tt (* m n) = (* (order h) (order k))}, both equalities must hold:
\begin{small}
\begin{verbatim}
  (defthmd rel-prime-factors-orders
    (implies (and (groupp g) (abelianp g)
                  (posp m) (posp n) (= (gcd m n) 1)
                  (= (order g) (* m n)))
             (let ((h (subgroup-ord-dividing m g)) (k (subgroup-ord-dividing n g)))
               (and (equal (order h) m) (equal (order k) n)))))
\end{verbatim}
\end{small}

Let {\tt p} be the least prime divisor of {\tt (order g)}.  Let {\tt m} be the maximum power of {\tt p} that divides {\tt (order g)} and let {\tt n = (/ (order g) m)}.  Then {\tt m} and {\tt n} are relatively prime.  We define a list of subgroups of {\tt g} recursively, using {\tt cyclic-p-subgroup-list}:
\begin{small}
\begin{verbatim}
  (defun cyclic-subgroup-list (g)
    (declare (xargs :measure (order g) ))
    (if (and (groupp g) (abelianp g))
        (if (= (order g) 1)
            ()
          (let* ((p (least-prime-divisor (order g)))
                 (m (max-power-dividing p (order g)))
                 (n (/ (order g) m))
                 (h (subgroup-ord-dividing m g))
                 (k (subgroup-ord-dividing n g)))
            (append (cyclic-p-subgroup-list p h)
                    (cyclic-subgroup-list k))))
      ()))
\end{verbatim}
\end{small}
The following is proved by induction, combining {\tt rel-prime-factors-intersection} and {\tt rel-prime\-factors-orders} with {\tt internal-direct-product-append} (Section~\ref{products}) and {\tt p-group-factorization} (Section~\ref{pgroups}):
\begin{small}
\begin{verbatim}
  (defthmd idp-cyclic-subgroup-list
    (implies (and (groupp g) (abelianp g) (> (order g) 1))
             (let ((l (cyclic-subgroup-list g)))
               (and (cyclic-p-group-list-p l)
                    (internal-direct-product-p l g)
                    (equal (product-orders l) (order g))))))
\end{verbatim}
\end{small}
Finally, we invoke isomorphismp-dp-idp (Section~\ref{products}):
\begin{small}
\begin{verbatim}
  (defthmd abelian-factorization
    (implies (and (groupp g) (abelianp g) (> (order g) 1))
             (let ((l (cyclic-subgroup-list g)))
               (and (cyclic-p-group-list-p l)
                    (isomorphismp (product-list-map l g) (direct-product l) g)))))
\end{verbatim}
\end{small}

\section{Uniqueness of the Factorization}\label{uniqueness}

We define the list of orders of a list of groups:
\begin{small}
\begin{verbatim}
  (defun orders (l)
    (if (consp l)
        (cons (order (car l)) (orders (cdr l)))
      ()))
\end{verbatim}
\end{small}
Our objective is to show that if the direct products of two lists of cyclic p-groups {\tt l} and {\tt m} are isomorphic, then {\tt (orders l)} and {\tt (orders m)} satisfy the following predicate, which is defined in the book {\tt lists}:
\begin{small}
\begin{verbatim}
  (defun permutationp (l m)
    (if (consp l)
        (and (member-equal (car l) m)
             (permutationp (cdr l) (remove1-equal (car l) m)))
      (endp m)))
\end{verbatim}
\end{small}
We shall make use of an equuivalent formulation of {\tt permutationp}, based on a function that counts the number of occurrences of an object in a list:
\begin{small}
\begin{verbatim}
  (defun hits (x l)
    (if (consp l)
        (if (equal x (car l))
            (1+ (hits x (cdr l)))
          (hits x (cdr l)))
      0))
\end{verbatim}
\end{small}
It is evident that {\tt (permutationp l m)} holds iff {\tt (hits x l) = (hits x m)} for all {\tt x}.  The formalization of this claim requires a function that searches for a counterexample:
\begin{small}
\begin{verbatim}
  (defun hits-diff-aux (test l m)
    (if (consp test)
        (if (equal (hits (car test) l) (hits (car test) m))
            (hits-diff-aux (cdr test) l m)
          (list (car test)))
      ()))
  (defund hits-diff (l m) (hits-diff-aux (append l m) l m))
  (defthmd hits-diff-perm (iff (permutationp l m) (not (hits-diff l m))))
\end{verbatim}
\end{small}

The uniqueness proof is also based on the notion of a power of an abelian group.  We define the list of {\tt n}th powers of the elements of g:
\begin{small}
\begin{verbatim}
  (defun power-list-aux (l n g)
    (if (consp l)
        (insert (power (car l) n g)
                (power-list-aux (cdr l) n g)
                g)
      ()))
  (defun power-list (n g)
    (power-list-aux (elts g) n g))
\end{verbatim}
\end{small}
If g is abelian, then this list forms a subgroup of g:
\begin{small}
\begin{verbatim}
  (defsubgroup group-power (n) g
    (and (posp n) (groupp g) (abelianp g))
    (power-list n g))
\end{verbatim}
\end{small}
If two abelian groups are isomorphic, then so are their {\tt n}th powers:
\begin{small}
\begin{verbatim}
  (defthmd isomorphismp-power
    (implies (and (isomorphismp map g h) (abelianp g) (posp n))
             (isomorphismp map (group-power n g) (group-power n h))))
\end{verbatim}
\end{small}
The {\tt n}th power of a direct product of abelian groups is the direct product of the {\tt n}th powers.  The proof is more challenging than expected, as it requires showing not only that each element list is a sublist of the other, but also that both lists are ordered with respect to {\tt (direct-product l)}, according to the lemma {\tt ind-dp-compare} (Section~\ref{products}):
\begin{small}
\begin{verbatim}
  (defun group-power-list (n l)
    (if (consp l)
        (cons (group-power n (car l))
              (group-power-list n (cdr l)))
      ()))
  (defthmd group-power-dp
    (implies (and (posp n) (consp l) (abelian-list-p l))
             (equal (group-power n (direct-product l))
                    (direct-product (group-power-list n l)))))
\end{verbatim}
\end{small}
The {\tt n}th power of a cyclic group is cyclic.  For prime {\tt p}, the order of {\tt (group-power p g)} depends on whether {\tt p} divides the order of {\tt g}:
\begin{small}
\begin{verbatim}
  (defun reduce-order (n p)
    (if (divides p n) (/ n p) n))
  (defthmd prime-power-cyclic
    (implies (and (cyclicp g) (primep p))
             (and (cyclicp (group-power p g))
                  (equal (order (group-power p g))
                         (reduce-order (order g) p)))))
\end{verbatim}
\end{small}                  
The list of orders of {\tt (group-power-list p l)}:
\begin{small}
\begin{verbatim}
  (defun reduce-orders (orders p)
    (if (consp orders)
        (cons (reduce-order (car orders) p) (reduce-orders (cdr orders) p))
      ()))
  (defthm order-group-power-list
    (implies (and (primep p) (cyclic-p-group-list-p l))
             (equal (orders (group-power-list p l))
                    (reduce-orders (orders l) p))))
\end{verbatim}
\end{small}

It is a simple matter to identify a prime that divides at least one of the orders:
\begin{small}
\begin{verbatim}
  (defund first-prime (l) (least-prime-divisor (order (car l))))
\end{verbatim}
\end{small}
Let {\tt p = (first-prime l)}.  We would like to use an induction scheme that replaces {\tt l} and {\tt m} with {\tt (group-powers p l)} and {\tt (group-powers p m)}, but in order to ensure that these lists inherit the properties of {\tt l} and {\tt m}, we must delete any occurrences of trivial groups:
\begin{small}
\begin{verbatim}
  (defun delete-trivial (l)
    (if (consp l)
        (if (= (order (car l)) 1)
            (delete-trivial (cdr l))
          (cons (car l) (delete-trivial (cdr l))))
      ()))
  (defund reduce-cyclic (l p) (delete-trivial (group-power-list p l)))
\end{verbatim}
\end{small}
Let {\tt l' = (reduce-cyclic l p)} and {\tt m' = (reduce-cyclic m p)}.  The properties of l and m are inherited by {\tt l'} and {\tt m'}:
\begin{small}
\begin{verbatim}
  (defthmd reduce-cyclic-p-group-list
    (implies (and (primep p) (cyclic-p-group-list-p l))
             (cyclic-p-group-list-p (reduce-cyclic l p))))
\end{verbatim}
\end{small}

We would like to show that if {\tt (orders l')} is a permutation of {\tt (orders m')}, then the same is true of {\tt l} and {\tt m}.  By {\tt hits-diff-perm}, it suffices to show that for all {\tt x}, {\tt (hits x (orders l)) = (hits x (orders m))}.  It may be proved as a consequence of {\tt order-group-power-list} that this holds for all {\tt x} other than {\tt p}.  But note that {\tt (orders l)} and {\tt (orders m)} have the same product:
\begin{small}
\begin{verbatim}
    (product-orders l) = (order (direct-product l))
                       = (order (direct-product l))
                       = (product-orders m).
\end{verbatim}
\end{small}
It follows that the equation holds for {\tt x = p} as well.  Thus, we have
\begin{small}
\begin{verbatim}
  (defthmd permutationp-orders
    (implies (and (consp l)
                (consp m)
                (cyclic-p-group-list-p l)
                (cyclic-p-group-list-p m)
                (primep p)
                (isomorphismp map (direct-product l) (direct-product m))
                (permutationp (orders (reduce-cyclic l p))
                              (orders (reduce-cyclic m p))))
           (permutationp (orders l) (orders m))))
\end{verbatim}
\end{small}

The base case of the induction is {\tt (or (null l') (null m'))}.  If {\tt l' = nil}, then every element of {\tt l} must be a group of order {\tt p}, which then every non-trivial element of {\tt (direct-product l)} has order {\tt p}.  But then the same must be true of {\tt (direct-product m)} and consequently m' = nil.  Therefore, if either l' or m' is null, then
\begin{small}
\begin{verbatim}
    (orders (reduce-cyclic l p)) = (orders (reduce-cyclic l p)) = nil.
\end{verbatim}
\end{small}
In particular, {\tt (permutationp (orders l') (orders m'))} and we have the following consequence of {\tt permutationp-orders}:
\begin{small}
\begin{verbatim}
  (defthmd null-reduce-cyclic-case
    (implies (and (consp l)
                  (consp m)
                  (cyclic-p-group-list-p l)
                  (cyclic-p-group-list-p m)
                  (primep p)
                  (or (null (reduce-cyclic l p))
                      (null (reduce-cyclic m p)))
                  (isomorphismp map (direct-product l) (direct-product m)))
             (permutationp (orders l) (orders m))))
\end{verbatim}
\end{small}

In the remaining inductive case, we need only show that if {\tt (direct-product l)} and {\tt (direct\-product m)} are isomorphic, then so are {\tt (direct-product l')} and {\tt (direct-product m')}.  We begin by constructing an isomorphism between {\tt (direct-product (group-power-list p l))} and {\tt (direct-prod\-uct l')}:
\begin{small}
\begin{verbatim}
  (defun delete-trivial-elt (x l)
    (if (consp x)
        (if (= (order (car l)) 1)
            (delete-trivial-elt (cdr x) (cdr l))
          (cons (car x) (delete-trivial-elt (cdr x) (cdr l))))
      ()))
  (defmap delete-trivial-iso (l)
    (group-tuples l)
    (delete-trivial-elt x l))
  (defthmd isomorphismp-delete-trivial
    (implies (and (group-list-p l) (consp (delete-trivial l)))
             (isomorphismp (delete-trivial-iso l)
                            (direct-product l)
                            (direct-product (delete-trivial l)))))
\end{verbatim}
\end{small}
Now suppose {\tt map} is an isomorphism from {\tt (direct-product l)} to {\tt (direct-product m)}.  By {\tt iso\-morphismp-power},
\begin{small}
\begin{verbatim}
   (isomorphismp map (group-power p (direct-product l))
                     (group-power p (direct-product m))),
\end{verbatim}
\end{small}
and by {\tt group-power-dp},
\begin{small}
\begin{verbatim}
   (isomorphismp map (direct-product (group-power-list p l))
                     (direct-product (group-power-list p m))).
\end{verbatim}
\end{small}
Thus, the desired isomorphism is constructed as a composition of three isomorphisms:
\begin{small}
\begin{verbatim}
  (defund reduce-cyclic-iso (map l m p)
    (compose-maps
      (delete-trivial-iso (group-power-list p m))
      (compose-maps
        map
        (inv-isomorphism (delete-trivial-iso (group-power-list p l))
                         (direct-product (group-power-list p l))
                         (direct-product (reduce-cyclic l p))))))
\end{verbatim}
\end{small}
We apply {\tt isomorphismp-delete-trivial}, {\tt isomorphismp-inv}, and {\tt isomorphismp-compose-maps} to conclude that {\tt reduce-cyclic-iso} is an isomorphism:
\begin{small}
\begin{verbatim}
  (defthmd isomorphismp-reduce-cyclic
    (implies (and (consp l)
                  (consp m)
                  (primep p)
                  (cyclic-p-group-list-p l)
                  (cyclic-p-group-list-p m)
                  (consp (reduce-cyclic l p))
                  (consp (reduce-cyclic m p))
                  (isomorphismp map (direct-product l) (direct-product m)))
             (isomorphismp (reduce-cyclic-iso map l m p)
                           (direct-product (reduce-cyclic l p))
                           (direct-product (reduce-cyclic m p)))))
\end{verbatim}
\end{small}
Our theorem follows from {\tt null-reduce-cyclic-case}, {\tt permutationp-orders}, and {\tt isomorphismp\-reduce-cyclic} by induction:
\begin{small}
\begin{verbatim}
  (defthmd abelian-factorization-unique
    (implies (and (consp l)
                  (consp m)
                  (cyclic-p-group-list-p l)
                  (cyclic-p-group-list-p m)
                  (isomorphismp map (direct-product l) (direct-product m)))
             (permutationp (orders l) (orders m))))
\end{verbatim}
\end{small}

\nocite{*}
\bibstyle{eptcs}
\bibliographystyle{eptcs}
\bibliography{groups2}
\end{document}